\documentclass[%
 reprint,
 amsmath,amssymb,
 aps,
 prl,
 longbibliography,
 lengthcheck,%
]{revtex4-1}

\usepackage{graphicx}
\usepackage{dcolumn}
\usepackage{bm}
\usepackage{hyperref}

\begin{document}

\preprint{APS/123-QED}

\title{Bright-like soliton solution in quasi-one-dimensional BEC in third order on interaction
radius
}

\author{P. A. Andreev}%
\email{andreevpa@physics.msu.ru}
\affiliation{
 Department of General Physics,
 Faculty of Physics, Moscow State
University, Moscow, Russian Federation.}
\author{L. S. Kuzmenkov}%
 \email{lsk@phys.msu.ru}
\affiliation{%
 Department of Theoretical Physics, Physics Faculty, Moscow State
University, Moscow, Russian Federation.}%




\date{\today}

\begin{abstract}
Nonlinear Schr\"{o}dinger equations and corresponding quantum
hydrodynamic (QHD) equations are widely used in studying ultracold
boson-fermion mixtures and superconductors. In this article, we
show that a more exact account of interaction in Bose-Einstein
condensate (BEC), in comparison with the Gross-Pitaevskii (GP)
approximation, leads to the existence of a new type of solitons.
We use a set of QHD equations in the third
order by the interaction radius (TOIR), which corresponds to the
GP equation in a first order by the interaction radius. The
solution for the soliton in a form of expression for the particle
concentration is obtained analytically. The conditions of
existence of the soliton are studied. It is shown what solution
exists if the interaction between the particles is repulsive. Particle concentration of order of $10^{12}$-$10^{14}$ $cm^{-3}$ has
been achieved experimentally for the BEC, the solution exists if
the scattering length is of the order of 1 $\mu$m, which can be
reached using the Feshbach resonance. It is one of the limit case
of existence of new solution. The corresponding scattering length
decrease with the increasing of concentration of particles. The
investigation of effects in the TOIR approximation gives a more detail
information on interaction potentials between the atoms and can be
used for a more detail investigation into the potential structure.
\end{abstract}

\maketitle


\section{\label{sec:level1}I. Introduction}

Nonlinear structures, solitons, and vortices have been actively
studied in atomic Bose-Einstein condensate (BEC) and boson-fermion
mixtures situated in magnetic and optical traps ~\cite{Young-S JP
B 11}-~\cite{Wen JP B 10}. Solitons exist in inquest of
nonlinearity of interactionally conditioned terms. These nonlinear
terms compensate the dispersion emerging particularly as a
consequence of a free motion of quantum particles. Solitons in the
BEC are described via the Gross-Pitaevskii (GP) equation
~\cite{L.P.Pitaevskii RMP 99} which has form of one-particle
nonlinear Schr\"{o}dinger equation (NLSE). The NLSE plays an
important role when describing the dynamics of various physical
systems; name just a few degenerated chargeless bosons and
fermions as well as superconductors. A more detailed account of
the interaction in comparison with the GP equation leads to the
appearance of additional terms both in the GP equation and in the
corresponding quantum hydrodynamics (QHD) equations. A similar
generalization is acquired for ultracold boson-fermion mixtures
~\cite{Andreev PRA08}. Such generalization leads to that the term
depending on the spatial derivatives of the concentration (modulus
quadrate of wave-function in the medium) appears in the NLSE.
Different authors ~\cite{Rosanov}, ~\cite{Braaten} suggested the
NLSEs for the description of the BEC dynamic taking into account
the terms depending on high degrees of concentration. The
occurrence of additional terms in the GP equation leads to varying
the characteristics of wave perturbations ~\cite{Andreev
Izv.Vuzov. 09 1}, solitons ~\cite{Andreev Izv.Vuzov. 10},
vortices, dispersion shock waves, and can also lead to new types
of solutions. The last point is considered in this paper.

The possibility to derive the GP equation from a microscopic
many-particle Schr\"{o}dinger equation (MPSE) is substantiated in
~\cite{Erdos PRL 07}. The method of direct derivation of the GP
equation from the MPSE was suggested in Ref. ~\cite{Andreev
PRA08}. It was made by means of QHD method ~\cite{Maksimov QHM
99-01}, ~\cite{Andreev PRB 11}. It is well-known that the GP
equation can be present in the form of hydrodynamic equations
~\cite{L.P.Pitaevskii RMP 99}
\begin{equation}\label{TOIR Sol cont from GP}\partial_{t}n(\textbf{r},t)+\partial^{\alpha}(n(\textbf{r},t)v^{\alpha}(\textbf{r},t))=0\end{equation}
and
$$mn(\textbf{r},t)\partial_{t}v^{\alpha}(\textbf{r},t)+\frac{1}{2}mn(\textbf{r},t)\partial^{\alpha}v^{2}(\textbf{r},t)$$
$$ -\frac{\hbar^{2}}{4m}\partial^{\alpha}\triangle n(\textbf{r},t)+\frac{\hbar^{2}}{4m}\partial^{\beta}\biggl(\frac{\partial^{\alpha}n(\textbf{r},t)\cdot\partial^{\beta}n(\textbf{r},t)}{n(\textbf{r},t)}\biggr)$$
\begin{equation}\label{TOIR Sol eiler from GP}+gn(\textbf{r},t)\partial^{\alpha}n(\textbf{r},t)=-n(\textbf{r},t)\partial^{\alpha}V_{ext}(\textbf{r},t),\end{equation}
where
$$g=\int d\textbf{r}U(r),$$
and $n(\textbf{r},t)$ is the concentration of particles and
$v^{\alpha}(\textbf{r},t)$ is the velocity field. The quantity
$\triangle$ is the Laplace operator.

For dilute gases the quantity $g$ can be express via scattering
length by formula
$$g=\frac{4\pi\hbar^{2}a}{m},$$
where $a$ is the scattering length.

In this paper we use set of equations derived with the QHD method.
There are different methods for obtaining equations describing the
BEC evolution. For example, in Ref. ~\cite{Nakamura AnnPh 11},
equation for BEC evolution was derived in the framework of
nonequilibrium Thermo Field Dynamics. In the QHD method the system
of equation is appeared directly from many-particle
Schr\"{o}dinger equation. The first step of derivation is the
definition of concentration of particles in three dimensional
physical space. Differentiation of concentration with respect to
time and applying of the Schr\"{o}dinger equation leads to
continuity equation, a current of density is arisen there. Next
step of derivation is differentiation of the current density. In
this way we obtain a momentum balance equation, in another terms
the Euler equation. A force field exists in the obtained Euler
equation. For neutral particles with the short-range interaction
the force field could be present in the form of a expansion in a
series. In this case, the GP equation emerges when we take into
account the first member of decomposition of the force field by
the interaction radius. The next nonzero term appears in the third
order by the interaction radius (TOIR).

Different types of solitons occur in the BEC and boson-fermion
mixtures. If the interaction between the bosons is repulsive
$a>0$, dark solitons that are the regions with a lowered
concentration of the particles can propagate in the BEC
~\cite{Burger et.al.}, ~\cite{Denschlag Sc 00}, ~\cite{Anderson
PRL 01}. Bright solitons, i.e., solitons of compression, can exist
in the system of Bose particles coupled by attractive forces in
quasi-one-dimensional (1D) traps ~\cite{Khaykovich et.al.},
~\cite{Strecker Nat 02}. Gap solitons manifest themselves in
periodic structures, particularly, the existens of gap bright
solitons are found experimentally in the system of bosons with
$a>0$ ~\cite{Eiermann cond-mat}. Solitons of compression occur in
boson-fermion mixtures if repulsive forces $a_{bb}>0$ act between
the bosons, while the interaction between bosons and fermions is
attractive with force $a_{bf}<0$ ~\cite{Karpiuk PRL 04}. In the
work ~\cite{Andreev Izv.Vuzov. 10} authors obtain a change of
the form of the well-known bright soliton due to TOIR terms. The bright
soliton solution arise from GP equation. More detailed account of
interaction with accuracy to TOIR leads to change of form of
bright soliton.

In this paper, we report about the existence of a new soliton
solution in a one dimensional (1D) BEC. This solution appears when we
account the interaction accurate to the TOIR. To obtain this
soliton we solve the set of the QHD equations by the perturbative
method suggested Washimi et al. ~\cite{Washimi PRL 66}, which is
widely used in the plasma physics, see for example ~\cite{perturb
meth}, ~\cite{Infeld book}. The obtained solution is the soliton
of compression; it exists under the condition $a>0$, i.e., in the
case of repulsion between the particles. The existence of such a
solution can be conditioned by higher spatial concentration
derivatives in the term for interaction in the TOIR. We consider a
two cases it are 1D configuration and quasi-1D trap. Let us notice
the limiting cases of existence of the solution. One of the
limiting cases in the region of parameters when the scattering
length (SL) $a$ is of the order $10^{-8} cm$, and the
corresponding equilibrium concentration is $10^{18} cm^{-3}$. That
could be actual in connection with the development of cooling
methods for dense gases ~\cite{Vogl Nat 09}, ~\cite{Sheik NatP
09}. Another limiting case is the region of parameters with the SL
of the order $10^{-4} cm$. This case corresponds to concentrations
 about of $10^{12}-10^{14} cm^{-3}$, which are usually dealt with
in experiments with BEC. Thus, in order to form the conditions for
soliton occurrence, the Feshbach resonance (FR) phenomenon should
be used ~\cite{Chin RMP 10}, ~\cite{Bloch RMP 08}. They attains
the wide-limit SL change in FR experiments, particularly the
values $10^{3}-10^{4} a_{0}$ ($a_{0}$-Bohr radius) can be reached
for magnetically trapped $^{85}Rb$ ~\cite{Cornish PRL 00}.

We use in this paper short-range interaction potential quantum
hydrodynamic equation derived for the system of ultracold neutral particles. In
connection with this, our attention should be paid to the fact
that an increase in the SL can be caused both by a decrease and an
increase in the depth or width of the interaction potential.
Assuming that an increase in the SL is caused by a decrease in the
interaction potential depth, the conditions of existence of
equations could be considered as fulfilled.

In a general case, the fact that under the FR condition larger
values of SL $a$ are attained,  can point to the fact that a more
successive account for the interaction should be necessary.

The processes and effects in the TOIR, along with the effects in
the spinor BEC ~\cite{Szankowski PRL 10}, magnetically
~\cite{Gligoric PRA 10}, ~\cite{Wilson PRL 10}, ~\cite{Bijnen PRA
10} and electrical ~\cite{Fischer PRA 06}, ~\cite{Ticknor PRL 11}
polarized BEC, can play an important role when investigating BEC
and interatomic interaction.

Our paper is organized as follows. In Sect. 2 we present basic
equation and describe using model. In Sect. 3 we consider a
solitons in 1D BEC and describe a method of getting of solution.
We show that with solution is a new solution and receive a
condition of existence of this solution. In Sect. 4 we obtain
system of QHD equations for the quasi-one dimensional case. In
Sect. 5 we investigate soliton solution obtained in sect. 3 for
quasi-one dimensional case. In Sect. 6 brief summary of obtained
results is presented.

\section{\label{sec:level1} II. Model}

To investigate solitons in BEC, we use the set of QHD equations up
to the TOIR approximation ~\cite{Andreev PRA08}. The calculation
of the first member in a quantum stress tensor that corresponds to
the GP equation is fulfilled in ~\cite{Andreev PRA08} under the
condition that the particles do not interact. A more complete
investigation into the conditions of derivation of the GP equation
from the MPSE shows that the GP equation appears in the first
order by the interaction radius (FOIR), if the particles are in an
arbitrary state that can be simulated by a single-particle wave
function. Such a state can particularly appears as a result of
strong interaction between the particles that takes place in
quantum fluids.

The QHD equations set for the atoms with a two-particle
interaction with the potential $U(r)$ and located in external
field $V_{ext}(\textbf{r},t)$ in the TOIR approximation has the
form ~\cite{Andreev PRA08}
\begin{equation}\label{TOIR Sol cont boze TOIR}\partial_{t}n(\textbf{r},t)+\partial^{\alpha}(n(\textbf{r},t)v^{\alpha}(\textbf{r},t))=0\end{equation}
and
$$mn(\textbf{r},t)\partial_{t}v^{\alpha}(\textbf{r},t)+\frac{1}{2}mn(\textbf{r},t)\partial^{\alpha}v^{2}(\textbf{r},t)$$
$$ -\frac{\hbar^{2}}{4m}\partial^{\alpha}\triangle n(\textbf{r},t)+\frac{\hbar^{2}}{4m}\partial^{\beta}\biggl(\frac{\partial^{\alpha}n(\textbf{r},t)\cdot\partial^{\beta}n(\textbf{r},t)}{n(\textbf{r},t)}\biggr)$$
$$ -\Upsilon n(\textbf{r},t)\partial_{\alpha}n(\textbf{r},t)-\frac{1}{16}\Upsilon_{2}\partial_{\alpha}\triangle n^{2}(\textbf{r},t)$$
 \begin{equation}\label{TOIR Sol eiler boze TOIR}=-n(\textbf{r},t)\partial^{\alpha}V_{ext}(\textbf{r},t),\end{equation}
where
\begin{equation}\label{TOIR Sol Upsilon} \Upsilon=\frac{4\pi}{3}\int
dr(r)^{3}\frac{\partial U(r)}{\partial r}
\end{equation}and
\begin{equation}\label{TOIR Sol Upsilon 2}\Upsilon_{2}=\frac{4\pi}{15}\int dr
(r)^{5}\frac{\partial U(r)}{\partial r}.\end{equation} We also
have $\Upsilon=-g$. Equations (\ref{TOIR Sol cont boze TOIR}) and (\ref{TOIR Sol eiler boze TOIR})
determine the dynamic of concentration of particles
$n(\textbf{r},t)$ and velocity field $v^{\alpha}(\textbf{r},t)$.
From equation (\ref{TOIR Sol eiler boze TOIR}) we see that
dynamics of BEC depends on different moments of interaction
potential $\Upsilon$, $\Upsilon_{2}$. The system of equations (\ref{TOIR Sol cont boze TOIR}) and
(\ref{TOIR Sol eiler boze TOIR}) is differ from (\ref{TOIR Sol cont from GP}) and (\ref{TOIR Sol
eiler from GP}) by existence of one new term. It is a last term in
left hand side of equation (\ref{TOIR Sol eiler boze TOIR}). This
term appears at interaction account up to TOIR approximation.

In diluted alkali gases, the interaction between particles can be considered
as scattering. In this case, the FOIR interaction constant can be
expressed in terms of SL $\Upsilon=-4\pi\hbar^{2}a/m$
~\cite{L.P.Pitaevskii RMP 99}. The second interaction constant $\Upsilon_{2}$
emerges in the TOIR. In a general case, parameter $\Upsilon_{2}$
is independent of $\Upsilon$. In ~\cite{Andreev PRA08}, the
approximate expression $\Upsilon_{2}$ via $\Upsilon$ is
considered. We use this expression in our work when we investigate
the existence region of the soliton solution.

Considering the dispersion equation for elementary excitations in
BEC accurate to TOIR
$$\omega^{2}(k)=\biggl(\frac{\hbar^{2}}{4m^{2}}+\frac{n_{0}\Upsilon_{2}}{8m}\biggr)k^{4}-\frac{\Upsilon n_{0}}{m_{}}k^{2}, $$
which obtained in ~\cite{Andreev PRA08} we can see that
coefficient at $k^{4}$ could be negative. It is realized at
condition $\Upsilon_{2}<-2\hbar^{2}/mn_{0}$. Consequently, we can
expect  that value $\Upsilon_{2}=-2\hbar^{2}/mn_{0}$ could play
important role at investigation of nonlinear processes.

\section{\label{sec:level1} III. Bright-like soliton in 1D BEC}

In this  section, we consider the solitons in the 1D BEC. For this
purpose, we use the perturbative method ~\cite{Washimi PRL 66},
~\cite{perturb meth}. Here we present some detail of calculations
and describe the perturbative method.

We investigate the case when the stretched variables include the
expansion parameter in follows combination:
\begin{equation}\label{TOIR masht of var 1D} \begin{array}{ccc}\xi=\varepsilon^{1/2}(z-ut),&\tau=\varepsilon^{3/2}ut &\end{array}\end{equation}
where $u$ is the phase velocity of the wave, $\varepsilon$- is a
small nondimension parameter.

An operational relations are arisen from (\ref{TOIR masht of var
1D})
\begin{equation}\label{TOIR repres. of var. 1D}\begin{array}{ccc} \partial_{x}=\varepsilon^{1/2}\partial_{\xi},&\partial_{t}=u\biggl(\varepsilon^{3/2}\partial_{\tau}-\varepsilon^{1/2}\partial_{\xi}\biggr)&\end{array}\end{equation}

The decomposition of the concentration and velocity field involves
a small parameter $\varepsilon$ in the following form:
\begin{equation} \label{TOIR Sol expansion 1D of conc}n=n_{0}+\varepsilon n_{1}
+\varepsilon^{2}n_{2}+...\end{equation}
\begin{equation} \label{TOIR Sol expansion 1D of vel}v=\varepsilon v_{1}
+\varepsilon^{2}v_{2}+...\end{equation} Presented in (\ref{TOIR
Sol expansion 1D of conc}) equilibrium concentration $n_{0}$ is a
constant. We put   expansions (\ref{TOIR repres. of var.
1D})-(\ref{TOIR Sol expansion 1D of vel}) in equations (\ref{TOIR Sol cont boze TOIR}) and (\ref{TOIR
Sol eiler boze TOIR}). Then, the system of equation is divided
into systems of equations in different orders on $\varepsilon$.

Equations emerging in the first order by $\varepsilon$ from the
system of equations (\ref{TOIR Sol eiler boze TOIR QODim}) have
form
$$-u\partial_{\xi}n_{1}+n_{0}\partial_{\xi}v_{1}=0,$$
\begin{equation} \label{TOIR Sol 1 order syst}-mun_{0}\partial_{\xi}v_{1}=\Upsilon n_{0}\partial_{\xi}n_{1}\end{equation}
and lead to the following expression for the phase velocity $u$:
\begin{equation} \label{TOIR Sol rel for U 1D}u^{2}=-\frac{\Upsilon n_{0}}{m}.\end{equation}
Square of phase velocity $u^{2}$ must be positive. Consequently
$\Upsilon$ is negative, i.e.
\begin{equation} \label{TOIR Sol cond of ex 1D}\Upsilon<0.\end{equation}
It corresponds to the repulsive SRI.

Also, from (\ref{TOIR Sol 1 order syst}) we obtain relation
between $n_{1}$ and $v_{1}$ and their derivatives
$$\partial_{\xi}n_{1}=\frac{n_{0}}{u}\partial_{\xi}v_{1}.$$
Integrating this equation and using a boundary conditions
\begin{equation}\label{TOIR boundari cond the first} \begin{array}{cccc} n_{1} ,& v_{1}\rightarrow 0 & at &
x\rightarrow\pm\infty\end{array}\end{equation} we have
\begin{equation}\label{TOIR 1e connection of var}n_{1}=\frac{n_{0}}{u}v_{1}.\end{equation}

In the second order by $\varepsilon$, from equations (\ref{TOIR Sol cont boze TOIR}) and (\ref{TOIR
Sol eiler boze TOIR}), we derive
\begin{equation}\label{TOIR 2e cont eq}-u\partial_{\xi}n_{2}+u\partial_{\tau}n_{1}+\partial_{\xi}(n_{0}v_{2}+n_{1}v_{1})=0\end{equation}
and
$$-mu(n_{0}\partial_{\xi}v_{2}+n_{1}\partial_{\xi}v_{1})+mun_{0}\partial_{\tau}v_{1}+mn_{0}v_{1}\partial_{\xi}v_{1}$$
\begin{equation} \label{TOIR Sol 2 order syst}-\frac{\hbar^{2}}{4m}\partial_{\xi}^{3}n_{1}=\Upsilon n_{0}\partial_{\xi}n_{2}+\Upsilon n_{1}\partial_{\xi}n_{1}+\frac{1}{8}\Upsilon_{2}n_{0}\partial_{\xi}^{3}n_{1}.\end{equation}

In (\ref{TOIR 2e cont eq}) we can express $n_{2}$ via $v_{2}$ and
$n_{1}$, $v_{1}$ \textit{and} put it in equation (\ref{TOIR Sol 2 order syst}). Using (\ref{TOIR Sol rel for U 1D}), we exclude $v_{2}$ from the
obtained equation
(\ref{TOIR Sol 2 order syst}). Thus, we obtain an equation which contain
$n_{1}$ and $v_{1}$, only. Using (\ref{TOIR 1e connection of
var}), expressing $v_{1}$ via $n_{1}$ we get a Korteweg-de Vries
equation for $n_{1}$
\begin{equation} \label{TOIR Sol KdV eq sos 1D}\partial_{\tau}n_{1}
+p_{1D}n_{1}\partial_{\xi}n_{1}+q_{1D}\partial_{\xi}^{3}n_{1}=0.\end{equation}
In this equation the coefficients $p_{1D}$ and $q_{1D}$ arise in
the form
\begin{equation} \label{TOIR Sol KdV koef 1D 1}p_{1D}=\frac{3}{2n_{0}},\end{equation}
and
\begin{equation} \label{TOIR Sol KdV koef 1D 2}q_{1D}=\frac{\frac{\hbar^{2}}{2m}+\frac{1}{8}n_{0}\Upsilon_{2}}{2n_{0}\Upsilon}.\end{equation}

From
equation (\ref{TOIR Sol KdV eq sos 1D}) we can find the solution in the form of a solitary wave using transformation
$\eta=\xi-V\tau$ \emph{and} taking into account boundary condition
$n_{1}=0$ and $\partial_{\eta}^{2}n_{1}=0$ at
$\eta\rightarrow\pm\infty$, we get
\begin{equation} \label{TOIR Sol solution of KdV 1D}n_{1}=\frac{3V}{p_{1D}}\frac{1}{\cosh^{2}\biggl(\frac{1}{2}\sqrt{\frac{V}{q_{1D}}}\eta\biggr)},\end{equation}
where $V$ is the velocity of solition propagation to the right.
From expression $p_{1D}=3/2n_{0}$ and solution (\ref{TOIR Sol
solution of KdV 1D}) we can find that a perturbation of
concentration is positive. Consequently, obtained solution is the
bright like soliton (BLS). A width of the soliton is given with
formula $d=2\sqrt{q_{1D}/V}$. BLS exists in the case $q_{1D}$ is
positive. From condition $q_{1D}>0$ (\ref{TOIR Sol cond of ex 1D})
we have
\begin{equation} \label{TOIR Sol solution existence cond}\frac{\hbar^{2}}{2m}+\frac{1}{8}n_{0}\Upsilon_{2}<0.\end{equation}
Relation (\ref{TOIR Sol solution existence cond}) is fulfil only
in the case when $\Upsilon_{2}$ is negative. In the absence of the
second interaction constant $\Upsilon_{2}$ (i. e. in the
Gross-Pitaevskii approximation) the relation (\ref{TOIR Sol
solution existence cond}) does not fulfil and, consequently, BLS
does not exist. From (\ref{TOIR Sol solution existence cond}) we
receive that the second interaction constant $\Upsilon_{2}$ must
be negative and it's module must be more than $4\hbar^{2}/mn_{0}$
\begin{equation} \label{TOIR Sol solution existence cond sv}|\Upsilon_{2}|>\frac{4\hbar^{2}}{mn_{0}}.\end{equation}

 Using representation $\Upsilon_{2}$ via the
s-wave SL $a$ ~\cite{Andreev PRA08} we get
\begin{equation} \label{TOIR Sol appr for G2}\Upsilon_{2}=\theta a^{2}\Upsilon=-4\pi\theta\hbar^{2}a^{3}/m ,\end{equation}
where $\theta$- is a constant, which is determined by an explicit form of
the interaction potential, $\theta>0$, $\theta\sim 1$ ~\cite{Andreev PRA08}.

From (\ref{TOIR Sol solution existence cond}) and (\ref{TOIR Sol
appr for G2}) we obtain
\begin{equation} \label{TOIR Sol solution existence cond SL}\pi a^{3}n_{0}>0.\end{equation}
It is the condition of BLS existence.

Due to used method perturbation $n_{1}$ must be smaller than
equilibrium concentration $n_{0}$: $n_{0}>>n_{1}$. Here we
consider the rate $n_{1}/n_{0}$ at the centre of soliton at
$\cosh(\sqrt{V/q_{1D}}/2\eta)=1$:
$$\frac{n_{1}(centre)}{n_{0}}=\frac{3V}{p_{1}n_{0}}=2V.$$
Correspondingly, dimensionless velocity $V$ must be much smaller
than one.

At equality in formula (\ref{TOIR Sol solution existence cond SL})
the BLS has infinite width, with the increasing of interaction
solitons width becomes finite. Formula (\ref{TOIR Sol solution
existence cond SL}) shows the bound condition for existence of
soliton.

Below we consider the same problem for cigar-shaped trap.

\section{\label{sec:level1}IV. The quantum hydrodynamics equation in the cigar-shaped traps}

Let us to consider the variation of the form of equations of QHD
in the case of cigar-shaped magnetic traps:
$$V_{ext}=\frac{m\omega_{0}^2}{2}(\rho^2+\lambda^2z^2),$$
where $\omega_{0}$ and $\lambda\omega_{0}$ are angular frequencies
in radial and axial directions and $\lambda$ is the anisotropy
parameter. In a quasi-1D geometry, the anisotropy parameter in the
axially-free motion approximation becomes zero $\lambda=0$. Thus,
the solution for the radial wave-function appears in the form
\begin{equation}\label{TOIR Sol Rad WF}\mid\Phi_{0}(\rho)\mid^{2}=n(\rho)=\frac{m\omega_{0}}{\pi\hbar}exp\biggl(-\frac{m\omega_{0}\rho^{2}}{\hbar}\biggr).\end{equation}

In the TOIR during the quasi-1D motion of bosons in magnetic
traps, the GP equation preserves the form but the interaction
constant ~\cite{Adhikari PRA05} changes. A complete 3D particle
concentration $n_{w}(\rho,z,t)$ can be presented as product of one-dimensional time dependent concentration $n(z,t)$ and static radial two-dimensional concentration $n(\rho)$
\begin{equation}\label{TOIR Sol concentr for rho and z}n_{w}(\rho,z,t)=n(z,t)n(\rho),\end{equation}
where the value $n(\rho)$ is presented by the formula (\ref{TOIR
Sol Rad WF}).

Applying the procedure described in ~\cite{Adhikari PRA05}, from
the set of equations (\ref{TOIR Sol eiler boze TOIR}) and using
the corresponding NLSE, we can acquire the system of QHD equations
for cigar-shaped trap. Here, we describe basic steps of this
procedure. Starting from equation (\ref{TOIR Sol eiler boze TOIR})
we get a equation of evolution of a following function, which sometimes called wave function in medium or order parameter,
$$\Phi(\textbf{r},t)=\sqrt{n(\textbf{r},t)}\exp(\imath m\theta(\textbf{r},t)/\hbar),$$
where $\theta$ is the potential of velocity field, i.e.
$\textbf{v}=\nabla\theta$. Equation for $\Phi(\textbf{r},t)$ is
the NLSE corresponding to system of equations (\ref{TOIR Sol eiler
boze TOIR}). Approximately we can present $\Phi(\textbf{r},t)$ in
the form $\Phi(\textbf{r},t)=\Phi(\rho,z,t)=\Phi(z,t)\Phi(\rho)$,
where $\Phi(\rho)$ is the wave function of the ground state of
harmonic oscillator and the square of module of $\Phi(\rho)$
presented by formula (\ref{TOIR Sol Rad WF}). Since, we get a NLSE
for $\Phi(z,t)$. This equation describes the evolution of BEC in
quasi-one dimensional trap. From obtained NLSE we derive the
system of QHD equations for quasi-1D trap. In the results we have
$$\partial_{t}n(z,t)+\partial_{z}(n(z,t)v(z,t))=0$$
and
$$mn(z,t)\partial_{t}v(z,t)+\frac{1}{2}mn(z,t)\partial_{z}v^{2}(z,t)$$
$$ -\frac{\hbar^{2}}{4m}\partial_{z}^{3}n(z,t)+\frac{\hbar^{2}}{4m}\partial_{z}\frac{(\partial_{z}n(z,t))^{2}}{n(z,t)}$$
 $$ +\alpha_{1} n(z,t)\partial_{z}n(z,t)+\alpha_{2} n(z,t)\partial_{z}^{3}n(z,t)$$
\begin{equation}\label{TOIR Sol eiler boze TOIR QODim}-\frac{7}{2}\alpha_{2}(\partial_{z} n(z,t))\partial_{z} n^{2}(z,t)
-\alpha_{2}\frac{(\partial_{z}n(z,t))^2}{n(z,t)}=0.\end{equation}

The following parameters appear in equation (\ref{TOIR Sol eiler
boze TOIR QODim}):
$$\alpha_{1}=-\Upsilon\frac{1}{2}\frac{m\omega_{0}}{\pi\hbar}+\frac{5\pi}{2}\Upsilon_{2}\Biggl(\frac{m\omega_{0}}{\pi\hbar}\Biggr)^{2}$$
and
$$\alpha_{2}=-\frac{3}{16}\Upsilon_{2}\frac{m\omega_{0}}{\pi\hbar}.$$

The form of nonlinear terms that describe the interaction in
equation (\ref{TOIR Sol eiler boze TOIR QODim}) differs from
corresponding terms in (\ref{TOIR Sol eiler boze TOIR}). This
leads to varying the form of solutions and conditions of their
existence in a quasi-1D geometry compared with a 1D case.

\section{\label{sec:level1} V. The small amplitude solitons in quasi-1D BEC}

In this  section, we consider the solitons in the BEC for the case
of small nonlinearity taking into account the TOIR. For this
purpose, we use the perturbative method ~\cite{Washimi PRL 66},
~\cite{perturb meth}.

We investigate the case when the stretched variables include the
expansion parameter in follows combination:
\begin{equation}\label{TOIR masht of var} \begin{array}{ccc}\xi=\varepsilon^{1/2}(z-ut),&\tau=\varepsilon^{3/2}ut &\end{array}\end{equation}
where $u$ is the phase velocity of  the wave.

The decomposition of the concentration and velocity field involves
a small parameter $\varepsilon$ in the following form:
\begin{equation} \label{TOIR Sol expansion of conc}n=n_{0}+\varepsilon n_{1}
+\varepsilon^{2}n_{2}+...\end{equation}
\begin{equation} \label{TOIR Sol expansion of vel}v=\varepsilon v_{1}
+\varepsilon^{2}v_{2}+...\end{equation}  Presented in (\ref{TOIR
Sol expansion of conc}) equilibrium concentration $n_{0}$ is a
constant. Equations emerging in the first order by $\varepsilon$
from the set of equations (\ref{TOIR Sol eiler boze TOIR QODim})
lead to the following expression for the phase velocity $u$:
$$u^{2}=\frac{n_{0}\alpha_{1}}{m}$$
\begin{equation} \label{TOIR Sol phase vel in sos}=\frac{n_{0}}{m}\biggl(-\Upsilon\frac{1}{2}\frac{m\omega_{0}}{\pi\hbar}+\frac{5\pi}{2}\Upsilon_{2}\biggl(\frac{m\omega_{0}}{\pi\hbar}\biggr)^{2}\biggr)\end{equation}
It is evident from (\ref{TOIR Sol phase vel in sos}) that the wave
can exist under the condition $\alpha_{1}>0$. The obtained
condition means that in the repulsive forces should act in the
case under consideration under the condition that the contribution
of  terms in FOIR prevails over the TOIR terms.

Relationship (\ref{TOIR Sol phase vel in sos}) is the analog of
the dispersion dependence. The use of scaling (\ref{TOIR masht of
var}) leads to simplifying the dispersion relationship compared
with the case when we consider small perturbations proportional to
$exp(-\imath\omega t+\imath kz)$. In the latter case, we obtained
the dispersion relation $\omega(k)$ form the set of equations
(\ref{TOIR Sol eiler boze TOIR QODim}) in the form:
$$\omega^{2}=\biggl(\frac{\hbar^{2}}{4m^{2}}
+\frac{3n_{0}}{16m}\Upsilon_{2}\frac{m\omega_{0}}{\pi\hbar}\biggr)k^{4}$$
\begin{equation}\label{TOIR disp rel gen QOD}
+n_{0}k^{2}\biggl(-\frac{1}{2}\Upsilon\frac{m\omega_{0}}{\pi\hbar}
+\frac{5\pi}{2}\Upsilon_{2}\biggl(\frac{m\omega_{0}}
{\pi\hbar}\biggr)^{2}\biggr).
\end{equation}
Thus, relationship (\ref{TOIR Sol phase vel in sos}) corresponds
to the phonon part of the dispersion dependence (\ref{TOIR disp
rel gen QOD}).

From the second-order set of equations (\ref{TOIR Sol eiler boze
TOIR QODim}) by $\varepsilon$ we find that the concentration
$n_{1}$ satisfies the Korteweg-de Vries equation:
\begin{equation} \label{TOIR Sol KdV eq sos}\partial_{\tau}n_{1}
+pn_{1}\partial_{\xi}n_{1}+q\partial_{\xi}^{3}n_{1}=0,\end{equation}
where
\begin{equation} \label{TOIR Sol p-q gen 1}p=\frac{3}{2n_{0}},\end{equation}
\begin{equation} \label{TOIR Sol p-q gen 2} q=\frac{4mn_{0}\alpha_{2}-\hbar^{2}}{8mn_{0}\alpha_{1}}.\end{equation}
When we derive this equation, we used boundary conditions $n_{1}=0$
and $v_{1}=0$ at $\xi\rightarrow\pm\infty$. Using transformation
$\eta=\xi-V\tau$, \textit{and}  taking into account boundary condition $n_{1}=0$
and $\partial_{\eta}^{2}n_{1}=0$ at $\eta\rightarrow\pm\infty$, we
can obtain the solution in the form of a solitary wave from
equation (\ref{TOIR Sol KdV eq sos})
\begin{equation} \label{TOIR Sol solution of KdV}n_{1}=\frac{3V}{p}\frac{1}{\cosh^{2}\biggl(\frac{1}{2}\sqrt{\frac{V}{q}}\eta\biggr)},\end{equation}
where $V$ is the velocity of solition propagation to the right.
Sign of perturbation is determined by the sign of $p$. From
formulas (\ref{TOIR Sol p-q gen 1}), (\ref{TOIR Sol p-q gen 2}) we
can see the quantity $p$ is positive. Consequently, obtained
solution is the soliton of compression or bright like soliton solution, by analogy with well-known bright soliton in BEC ~\cite{Andreev Izv.Vuzov. 10}, ~\cite{Khaykovich et.al.}, ~\cite{Strecker Nat 02}. As it will be shown
below, this solution exists only when taking into account the
TOIR.

Let us pass on to a detail consideration of the conditions of
existence of the solution (\ref{TOIR Sol solution of KdV}). The
solution (\ref{TOIR Sol solution of KdV}) of the equation
(\ref{TOIR Sol KdV eq sos}) exists as the conditions $q>0$ and
$\alpha_{1}>0$ are fulfilled. We start with consideration the condition $q>0$.
As $\alpha_{1}>0$, then to fulfill the condition $q>0$ we need
$-\hbar^{2}+4mn_{0}\alpha_{2}>0$.

 In the case when $\alpha_{2}$ is vanish (i.e. in FOIR approximation)
the solution (\ref{TOIR Sol solution of KdV}) is not exist. It
means, that solution arises in TOIR approximation which developed
in ~\cite{Andreev PRA08}. One of the condition of existence of the
solution (\ref{TOIR Sol solution of KdV}) is:
\begin{equation} \label{TOIR Sol cond of ex sol 1}\alpha_{2}>\frac{\hbar^{2}}{4mn_{0}}.\end{equation}
Consequently, for the second interaction constant $\Upsilon_{2}$ we
obtain:
\begin{equation} \label{TOIR Sol cond of ex sol 1f2}\Upsilon_{2}<-\frac{4\pi\hbar^{3}}{3m^{2}n_{0}\omega_{0}}.\end{equation}
Using relation (\ref{TOIR Sol appr for G2}) we can make estimation
for corresponding SL. It is useful to present the value of possible SL in
the terms of space parameter of the trap
$a_{\perp}=\sqrt{\hbar/m\omega_{0}}$. From conditions (\ref{TOIR
Sol cond of ex sol 1f2}), (\ref{TOIR Sol appr for G2}), the
conditions for SL $a$ appear:
\begin{equation} \label{TOIR Sol cond of ex sol 1f2 bbbb}a>\sqrt[3]{\frac{a_{\perp}^{2}}{3\theta n_{0}}}.\end{equation}
In addition, from $\alpha_{1}>0$ and (\ref{TOIR Sol appr for G2}),
we obtain
\begin{equation} \label{TOIR sec cond ex of sol}   a<\frac{a_{\perp}}{\sqrt{5\theta}}.\end{equation}
Using equation (\ref{TOIR Sol appr for G2}), the particle
concentration $n$ can be presented in the form
\begin{equation} \label{TOIR Sol wiev sol}n=n_{0}+2Vn_{0}\varepsilon\cdot sech^{2}\biggl(\frac{1}{2}\sqrt{\frac{V}{q'}}\eta\biggr),\end{equation}
where
\begin{equation} \label{TOIR Sol q'}q'=\frac{3\theta mn_{0}\omega_{0}a^{3}\hbar-\hbar^{2}}{16mn_{0}\omega_{0}a(\hbar-5a^{2}\theta m\omega_{0})}\end{equation}

\begin{figure}
\includegraphics[width=8cm,angle=0]{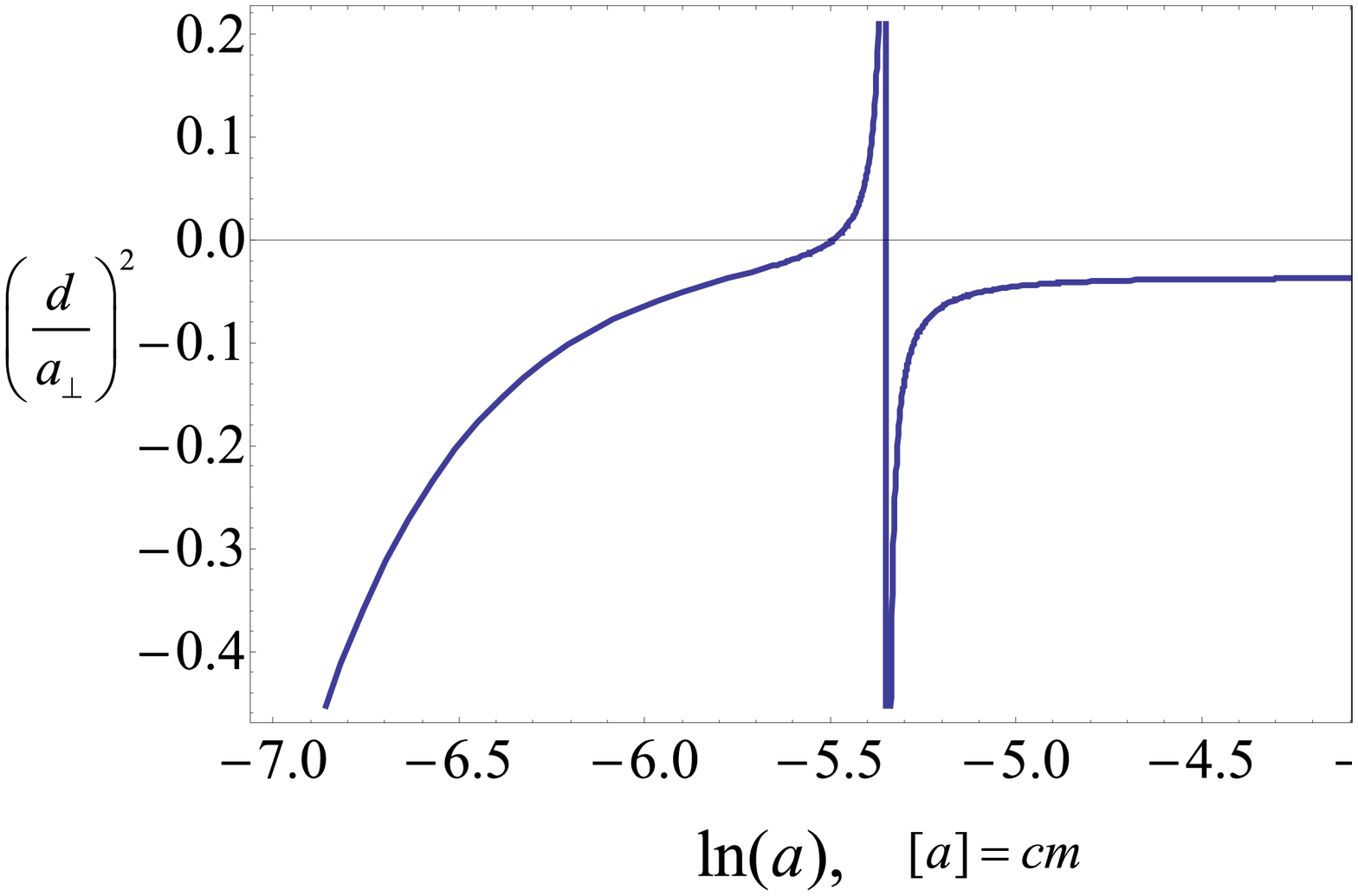}
\includegraphics[width=8cm,angle=0]{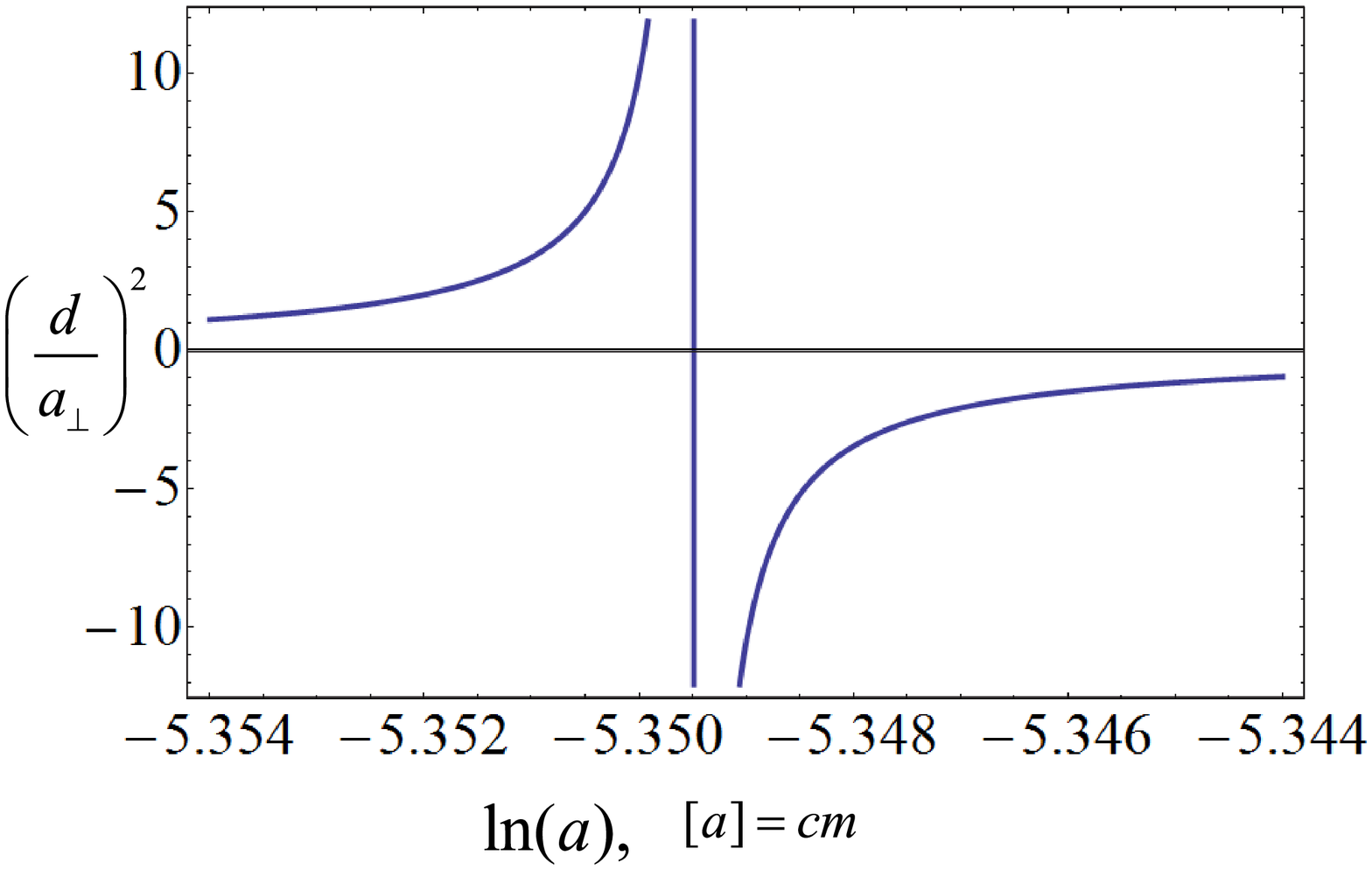}
\caption{\label{fig1:epsart} The dependence of the soliton width
$d$ on the scattering length $a$ at fixed parameter of the trap
$a_{\perp}=\sqrt{\hbar/m\omega_{0}}=10^{-5}cm$ and equilibrium
concentration $n_{0}=10^{6}cm^{-1}$ \textit{and} at $V=1$, $\theta=1$. On
Fig.1a we can see what width of soliton is the positive in
small range of the values of the SL. On Fig.1b the area of the
resonance is presented more detailed than on Fig.1a.}
\end{figure}

\begin{figure}
\includegraphics[width=8cm,angle=0]{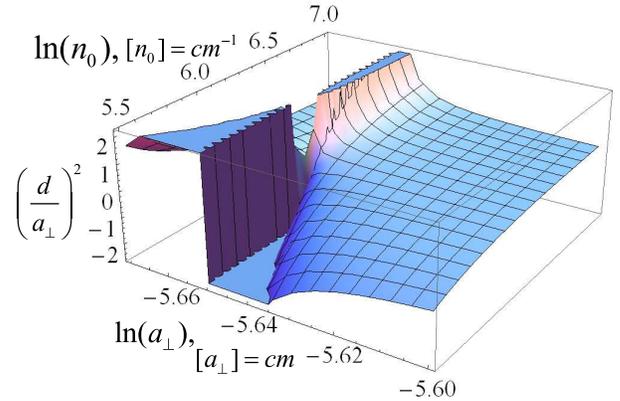}
\caption{\label{fig2:epsart} (Color online) The dependence of
soliton width $d$ on radial parameter of the trap $a_{\perp}$ and
the nonperturbative concentration of the particles $n_{0}$ at the
fixed scattering length $a=10^{-6}cm$, and at $V=1$, $\theta=1$.
The soliton width $d$ is positive in two area. But in area with
smaller value of the particles concentration $n_{0}$ the square of
phase velocity is negative. Thus, the solution exist in area of
bigger concentrations.}
\end{figure}

The soliton width $d$ arises in the form $d=2\sqrt{q'/V}$.  The
numerical analysis of formula (\ref{TOIR Sol q'}) is presented in
Fig. ~\ref{fig1:epsart}-~\ref{fig2:epsart}. It is evident from
Fig. ~\ref{fig1:epsart}-~\ref{fig2:epsart} that there is a narrow interval of the SL
values, for which the solution (\ref{TOIR Sol wiev sol}),
(\ref{TOIR Sol q'}) exists. The dependence of the soliton width on
the SL  is resonant-shaped. The resonant value of the
SL $a_{r}$ depends on the particle concentration $n_{0}$ and trap
parameter. The values of $a_{r}$ become lower at increasing of equilibrium concentration $n_{0}$. The
SL $a_{r}$ reaches 0.1 nm at the concentration of the order
$10^{18}$ $cm^{-3}$. As the concentration decreases to values
$10^{12}$-$10^{14}$ $cm^{-3}$, which are usually used in BEC
experiments, the SL increases to the values of the order of 1
$\mu$m. Such values of the SL can be attained when using the FR.

\section{\label{sec:level1} VI. Conclusion}

In this article, we showed that at a more exact accounting of the
interaction, specifically, taking into account the TOIR, a new
type of solitons emerges in the BEC. We also studied the
conditions of existence of such a solution. For this problem
solving we used the set of QHD equations where the interactions
included up to TOIR approximation. The TOIR approximation is an
example of the nonlocal interaction. The GP approximation gives us
the force density in the right hand side of the Euler equation
$\textbf{F}=-g\nabla n^{2}/2=\Upsilon\nabla n^{2}/2$. It is
corresponds to the first order on interaction radius. The
interaction including up to TOIR approximation gives us the second
tern in the force field $\textbf{F}=\Upsilon\nabla
n^{2}/2+\Upsilon_{2}\nabla\triangle n^{2}/16$. The new term
contain the third spatial derivative of the concentration square
and the new interaction constant. For obtained results analysis
and estimation we used approximate estimation of $\Upsilon_{2}$ \textit{and} 
its approximate connection with the $\Upsilon$ or SL $a$.

We found that BLS (soliton of compression) exists in 1D case (one
dimensional propagation in three dimensional medium) in the case of
strong enough repulsive interaction. BLS appearance strongly
connects with the second interaction constant $\Upsilon_{2}$. If we
consider the interaction in the first order by the interaction radius
there is no BLS. We also studied the BLS behavior in the case of
quasi-1D trap and describe contribution of external fields on BLS
amplitude and width.

In a general case, the second interaction constant $\Upsilon_{2}$,
which appears in the TOIR, is independent of $\Upsilon$ and,
consequently, of SL $a$. Thus, the relationships obtained in this
article (\ref{TOIR Sol p-q gen 1}), (\ref{TOIR Sol p-q gen 2}),
(\ref{TOIR Sol solution of KdV}) can be used for an independent
experimental determination of $\Upsilon_{2}$. In this case,
parameter $\hbar^{2}/mn_{w0}$ can be used for the qualitative
evaluation of the second interaction constant $\Upsilon_{2}$.

Thus, in this paper we showed that new physical effects appear at
account of interaction up to TOIR approximation. The processes and
effects in the TOIR approximation, along with the effects in the
spinor and polarized BEC, can play an important role at
investigation of BEC and interatomic interaction.

\end{document}